\documentclass[12pt]{article}
\usepackage{graphicx,amsmath,amssymb}
\catcode`\@=11
\long\def\@makefntext#1{
\protect\noindent \hbox to 3.2pt {\hskip-.9pt
$^{{\ninerm\@thefnmark}}$\hfil}#1\hfill}        %CAN BE USED

 \def\@makefnmark{\hbox to 0pt{$^{\@thefnmark}$\hss}}  %ORIGINAL

\def\ps@myheadings{\let\@mkboth\@gobbletwo
\def\@oddhead{\hbox{}
\rightmark\hfil\ninerm\thepage}
\def\@oddfoot{}\def\@evenhead{\ninerm\thepage\hfil
\leftmark\hbox{}}\def\@evenfoot{}
\def\sectionmark##1{}\def\subsectionmark##1{}}

\renewenvironment{thebibliography}[1]
    {\begin{list}{$^{\arabic{enumi}}$}
    {\usecounter{enumi}\setlength{\parsep}{0pt}
\setlength{\leftmargin 1.25cm}{\rightmargin 0pt}
     \setlength{\itemsep}{0pt} \settowidth
    {\labelwidth}{#1.}\sloppy}}{\end{list}}

\newcounter{itemlistc}
\newcounter{romanlistc}
\newcounter{alphlistc}
\newcounter{arabiclistc}

\def\@citex[#1]#2{\if@filesw\immediate\write\@auxout
    {\string\citation{#2}}\fi
\def\@citea{}\@cite{\@for\@citeb:=#2\do
    {\@citea\def\@citea{,}\@ifundefined
    {b@\@citeb}{{\bf ?}\@warning
    {Citation `\@citeb' on page \thepage \space undefined}}
    {\csname b@\@citeb\endcsname}}}{#1}}

\newif\if@cghi
\def\cite{\@cghitrue\@ifnextchar [{\@tempswatrue
    \@citex}{\@tempswafalse\@citex[]}}
\def\citelow{\@cghifalse\@ifnextchar [{\@tempswatrue
    \@citex}{\@tempswafalse\@citex[]}}
\def\@cite#1#2{{$\null^{#1}$\if@tempswa\typeout
    {IJCGA warning: optional citation argument
    ignored: `#2'} \fi}}

\def\fnt#1#2{\footnotetext{\kern-.3em
    {$^{\mbox{\sevenrm #1}}$}{#2}}}

 1
 1
 1
\font\ninerm=cmr9

\newcommand{\der}[2]{\frac{\partial #1}{\partial #2}}

\newcommand{\be}{\begin{equation}}
\newcommand{\ee}{\end{equation}}
\newcommand{\bea}{\begin{eqnarray}}
\newcommand{\eea}{\end{eqnarray}}
\newcommand{\gm}{\gamma}

\usepackage{subfigure}

\begin{document}
\title{Maximal slicings in spherical symmetry: \\ local existence and construction}
\author{Isabel Cordero-Carri\'on$^1$, Jos\'e Mar\'ia Ib\'a\~nez$^2$ \\ and Juan Antonio Morales-Lladosa$^3$}
%\date{\today}
\date{\empty}

\hyphenation{e-xis-ten-ce cha-rac-te-ris-tic con-tai-ning pa-ra-me-ter 
pre-vious pro-per-ty ins-tan-ce addi-tion rea-so-na-ble e-xa-mi-ned 
cons-truc-ted assess sa-tis-fying ge-ne-ral Eins-tein nu-me-ri-cal symme-try 
me-tric ma-xi-mal cons-traint pro-per-ly assu-res coe-ffi-cient using fa-mi-ly 
without di-ffe-ren-tia-ble a-na-lo-gous a-ccor-ding nu-me-ri-ca-lly va-lues 
leading gra-vi-ty pu-bli-shed fo-llo-wing dy-na-mi-cal sli-cings in-te-res-ting
cha-rac-te-ri-za-tion co-rres-pon-ding cons-tant sta-tio-na-ry written va-cuum
Gar-ching classi-cal}

\maketitle
\vspace*{-0.5cm}
\begin{abstract}

We show that any spherically symmetric spacetime locally admits a maximal spacelike slicing and we give a procedure allowing its construction. The construction procedure that we have designed is based on
purely geometrical arguments and, in practice, leads to solve a decoupled 
system of first order quasi-linear partial differential equations. We have 
explicitly built up maximal foliations in Minkowski and Friedmann spacetimes. 
Our approach admits further generalizations and efficient computational 
implementation. As by product, we suggest some applications of our work in the 
task of calibrating Numerical Relativity complex codes, usually written in 
Cartesian coordinates.

\end{abstract}

\vspace*{2mm}

\begin{center}
PACS: 04.20.Cv, 04.20.-q, 04.20.Ex, 04.25.D-, 02.30.Jr,
02.60.Lj
\end{center}

\vspace*{2cm} \noindent $^1$ Max-Planck-Institute for Astrophysics, Garching, 
Karl-Schwarschild-Str. 1, D-85741 Garching, Germany  E-mail: {\tt
chabela@mpa-garching.mpg.de}
\\
$^2$ Departament d'Astronomia i Astrof\'{\i}sica, Universitat de
Val\`encia, E-46100 Burjassot, Val\`encia, Spain.  E-mail: {\tt
Jose.M.Ibanez@uv.es}
\\
$^3$ Departament d'Astronomia i Astrof\'{\i}sica, Universitat de
Val\`encia, E-46100 Burjassot, Val\`encia, Spain.  E-mail: {\tt
antonio.morales@uv.es}
\newpage

\section{Introduction}
\label{sec:1}
In Lorentzian geometry, a {\it maximal\,} hypersurface is one that is spacelike and has
vanishing mean extrinsic curvature, $K = 0$, $K$ being the trace of the 
extrinsic curvature of the hypersurface. The name comes from the fact that the 
induced volume functional reaches a local maximum with respect the variations 
that keep fixed a given boundary. Maximal hypersurfaces were considered by 
Lichnerowicz \cite{Lichn44} to solve Einstein's constraint equations, giving 
motivation for subsequent studies on the subject.\cite{MmeChoquet76,CaFisMarsdMur76,MarsdenTip80,Bartnik84,ChrWald94,BurRen96}
In fact, the existence of maximal hypersurfaces is extensively used in 
Mathematical Relativity.\cite{MmeChoquet08} This property is a very simple 
geometric assumption allowing one to establish general results for a very large
variety of spacetimes, for instance, local or asymptotically stationary or 
conformally flat spacetimes.

In this paper, we will use the term {\it maximal slicing} when referring to a 
(non intersecting) family of spacelike maximal hypersurfaces which locally 
foliates a certain domain of spacetime. This type of slicing has very nice 
properties as, for example: i) the well-known singularity avoidance 
capability,\cite{SmarrY78b} ii) it is well adapted to the propagation of gravitational 
waves,\cite{SmarrY78a,ShibaN95} and, iii) it gives the natural Newtonian 
analogous when, in addition, conformal flatness is imposed on each slice.\cite{Isenb78} 
Maximal slicing condition has been recently used in the Fully 
Constrained Formulation of Einstein equations derived by the 
Meudon group,\cite{BonazGGN04,Lin-Novak06} in the field of Numerical 
Relativity.\cite{Gourg07a,Alcub08} The study of maximal hypersurfaces is 
not only interesting to carry out practical applications, but also, to solve 
fundamental questions in Relativity. Consider, for example, the proof of the 
positiveness of the mass for isolated gravitating systems, that was firstly 
proved by assuming the existence of a maximal spacelike hypersurface,\cite{SchShi79} 
and only then, it was finally established by removing this 
reasonable hypothesis.\cite{SchShi81}

The study of maximal slicings has been mainly motivated by the resolution of 
the initial value problem.\cite{Lichn44} This idea was popularized by Smarr 
and York.\cite{SmarrY78b,York79} Then, for a given class of spacetimes, the 
construction of a family of slices with a given property (e.g., maximal slicing
and/or conformal flatness) may be dynamically accomplished in an evolution 
scheme: assuming the existence of an initial 3-surface with that property, 
Einstein equations are used to evolve locally the corresponding additional 
constraint.

This strategy can be applied both analytically and numerically, as it has been 
done\ by Estabroook et al.\cite{Estabrook-WCWST73} and by Beig and Murchadha\cite{Beig-Mu98} to construct maximal slicings 
in the extended Schwarzschild geometry. This maximal foliation of the 
Kruskal-Szekeres black hole was obtained independently by Reinhart\cite{Reinhart73} from a different approach. Reinhart's procedure\cite{Reinhart73} does not make any use of Einstein equations, being 
essentially a geometrical (non dynamical) approach.

Similar examples of maximal slicings have been examined by Malec\cite{Malec94} in the 
Reissner-Nordstr$\ddot{\rm o}$m geometry, and by Petrich et al.\cite{PetriST85} in dynamical 
spherically symmetric spacetimes (SSSTs) of dust collapse scenarios, by using isotropic coordinates. Eardley and Smarr\cite{Eardley-Smarr79} obtained general results and presented some
conjectures about the existence of Cauchy time functions whose level surfaces 
are maximal (or foliations with constant mean extrinsic curvature) with direct applications to the 
Lema{\^{\i}}tre-Tolman-Bondi geometry. Of course, the study of the existence 
(and uniqueness) of other types of foliations deserves a lot of interest in 
SSSTs and more general spacetimes. That is the case of synchronizations by flat 
instants (Painlev\'e-Gullstrand slicings) and their generalizations.\cite{GuMur,HuQaSi,BeigSi,Doran,Natario,Lin,Alicia} Recently, 
foliations with constant mean curvature have been constructed by Malec\cite{Malec}
in the Schwarzschild geometry.

In spite of their extended use, the existence of maximal slicings in SSSTs has 
been only established for vacuum and for some particular energy 
contents.\cite{Isenb78,Estabrook-WCWST73,Beig-Mu98,Reinhart73,Malec94,PetriST85,Eardley-Smarr79}
There is, as far as we know, no theorem stating that always it is possible to 
build a maximal slicing in a SSST.

In this work, we aim to prove the local existence of maximal slicings in an 
arbitrary SSST. We will follow a purely geometrical approach, independent of
Einstein equations, complementary to the standard time evolution 
strategy.\cite{Isenb78,Estabrook-WCWST73,Beig-Mu98,Malec94,PetriST85,Eardley-Smarr79}
Although our study is independent of the field equations, one by-product of our
approach, which could be of interest in the field of Numerical Relativity, is 
that it provides a means to assess complex and sophisticated 3D numerical codes
built to solve Einstein equations. Indeed, most of the current 3D present-day 
Numerical Relativity codes are written in Cartesian coordinates. Two classical,
standard test-beds in spherical symmetry are the long-term stable evolution of 
a Schwarzschild black hole (vacuum), and the Oppenheimer-Snyder 
collapse\cite{OppSny} of dust (matter). In both cases the builder of a Numerical 
Relativity code aims at the recovering of the corresponding well-known 
solutions in spherical symmetry. This is an obvious assessment (in the 3D case)
of a code. Let us consider the Oppenheimer-Snyder analysis of the collapse of a
ball of dust; it is the simplest approach to the current studies on the complex
phenomena of the stellar core collapse (in the astrophysical scenario of 
hydrodynamical supernovae). Hence, for a given 3D numerical code, written in 
Cartesian coordinates, the Oppenheimer-Snyder solution is a nice test-bed to 
asses the performances of that code, but also to calibrate it against any 
spurious gravitational radiation released as a consequence of grid effects. 
Since our approach provides a procedure to generate a maximal slicing in a 
SSST, this, in practice, means that any SSST, not necessarily satisfying the 
maximal slicing condition, can be reformulated in terms of this slicing and 
used to check a numerical 3D solution generated by a code written in Cartesian 
coordinates or not and satisfying the maximal slicing gauge.

Our analysis has focused on SSSTs in standard coordinates. An intrinsic characterization of SSSTs in 
terms of metric tensorial concomitants has been very recently accomplished 
by Ferrando and S\'aez.\cite{Ferrando} The study of SSSTs is of particular interest in different 
domains of general relativity: classification of exact solutions of Einstein 
equations, simplified models of astrophysical and cosmological scenarios,\cite{PlebKra}  test-bed solutions for fully general-relativistic 
time-dependent numerical codes which evolve matter in strong gravitational 
fields.\cite{PetriST85,PetriST86,Eric92,Romero96} SSSTs provide 
the background for a perturbative theory. All these reasons advise us to start 
our analysis in such a simple symmetry and to defer in a future paper its 
extension to less strongly symmetric spacetimes.

The paper is organized as follows. In Sec.~\ref{sec:2} we consider (in a SSST)
vorticity-free and non-expanding observers, that is, those that are orthogonal 
to maximal slicings. In Sec.~\ref{sec:3} we analyze the existence of a 
coordinate transformation allowing a new coordinate time that has associated a 
maximal slicing. In Sec.~\ref{sec:4} we derive a decoupled system of three 
first order partial differential equations that proves the local existence of a
maximal slicing and, in Sec.~\ref{sec:5}, we provide a general procedure 
allowing its construction. This procedure is illustrated in Sec.~\ref{sec:6} 
considering the most simple geometry: a flat spacetime, solving the problem 
analytically. The procedure is also illustrated in Sec.~\ref{sec:7} considering
a closed Friedmann spacetime, which is used in physical applications of dust 
collapse. Finally, in Sec.~\ref{sec:8}, we summarize the results. It will be 
understood that the title of the sections always concern spacetimes with 
spherical symmetry. A short communication\cite{Isa-ere09} of preliminary results of this work 
was presented at the Spanish Relativity meeting ERE-2009.

%%%%%%%%%%%%%%%%%%%%%%%%%%%%%%%%%%%%%%%%

\section{Kinematical approach to maximal slicings}
\label{sec:2}
Without loss of generality, we start with the canonical form of the metric of a
SSST:
\be
    ds^2 = A\,dt^2 + 2C\,dt\ dr + B\,dr^2 + D\,d\Omega^2,
\label{e:sph_sym}
\ee
(see, e.g.  the book by Pleba{\'n}ski and Krasi{\'n}ski\cite{PlebKra}) where $d\Omega^2=d\theta^2 + \sin^2\theta d\varphi^2$ is the metric of the 
2-sphere, and $A, B, C, D$ are smooth functions of $t$ and $r$ such that 
$AB-C^2<0$, which guarantees the Lorentzian character of the metric. In 
addition we choose the signature $(-,+,+,+)$, and accordingly $D>0$. Partial 
derivatives with respect to $r$ and $t$ will be denoted as 
$\displaystyle{\der{f}{r} = f'}$ and $\displaystyle{\der{f}{t} = \dot{f}}$ 
respectively. The spatial metric $\gm_{ij}$ induced on the hypersurfaces 
$\Sigma_t$, defined by $t=\mathrm{constant}$, is 
$\gm_{ij} = \mathrm{diag}(B, D,D \, \sin^2\theta)$, where $B>0$ since we are 
considering spacelike hypersurfaces. Let $n = n^\mu \partial_\mu$ be the future
pointing timelike unit normal to the hypersurfaces $\Sigma_t$,
\be
    n = \frac{1}{\alpha}\left(\der{}{t} - \frac{C}{B}\der{}{r}\right), \;\;
\alpha = \sqrt{\frac{C^2}{B}-A}.
\label{e:def_n}
\ee
The mean extrinsic curvature $K$ of $\Sigma_t$ is
\be
    - K = \nabla_{\mu}n^{\mu} = \frac{1}{D \sqrt{C^2 - AB}} 
\Big[(\sqrt{B} D)\mbox{\huge ${^.}$} - \Big(\frac{CD}{\sqrt{B}}\Big)^{'}\Big],
\label{e:def_K}
\ee
where $\nabla$ is the covariant derivative with respect to the spacetime metric
given by Eq.~(\ref{e:sph_sym}), and $\mu \in \{t,r,\theta,\varphi\}$. Simplifying 
this relation, we obtain
\be
    K = \frac{1}{2\alpha B} \left( - \dot{B} - 2B\frac{\dot{D}}{D} + 2C'
- C\frac{B'}{B} + 2C\frac{D'}{D} \right).
\label{e:K}
\ee

That the surfaces $t=\mathrm{constant}$ are maximal means that the maximal 
slicing condition $K=0$ is fulfilled. In the following, we assume that 
$A, B, C$ and $D$ are known functions. We aim to prove the local existence of a
maximal slicing in spherical symmetry. From a kinematical point of view, we 
look for the existence of a non-expanding observer $u$ on the 2-surfaces 
defined by constant $\theta$ and $\varphi$, i.e., a unit and future directed 
radial timelike field $u = u^\mu = (u^t (t,r), u^r(t,r), 0, 0)$, such that 
$\nabla_{\mu} u^{\mu} = 0$ or, equivalently,
\be
     (u^t D \sqrt{C^2 - AB}){\mbox{\huge ${^.}$}} + (u^r D \sqrt{C^2 - AB})'=0.
\label{e:divergence-free}
\ee
The condition $g(u,u)=-1$, with respect to the spacetime metric $g_{\mu\nu}$ 
given by Eq.~(\ref{e:sph_sym}), leads to the following relation between the 
components of $u$:
\be
     A \, (u^t)^2 + B \, (u^r)^2 + 2 \, C \, u^t u^r = -1.
\label{e:unit}
\ee
Eq.~(\ref{e:unit}) allows to express $u^r$ in terms of $u^t$ and the metric 
components when the associated discriminant is not negative, that is,
\be
\label{ut}
	(u^t)^2 \geq \frac {B}{C^2-AB}.
\ee
Substituting the expression for $u^r$ in Eq.~(\ref{e:divergence-free}), one 
obtains a first order partial differential equation for $u^t$. In order to 
guarantee the existence of a maximal slicing, the solutions of this equation 
have to satisfy the above algebraic constraint~(\ref{ut}) on $u^t$.

In the next section, we introduce a change of coordinates 
$\left\{ \tilde{t} = \tilde{t}(t,r),\right.$ 
$\left. \tilde{r} = \tilde{r}(t,r), \theta, \varphi \right\}$ and we require 
that the new coordinate hypersurfaces  $\tilde{t} = \mathrm{constant}$ be 
maximal. As we show below, no additional algebraic constraint must be 
considered when the conditions assuring the existence of this coordinate 
transformation are properly taken into account.

%%%%%%%%%%%%%%%%%%%%%%%%%%%%%%%%%%%%%%%%%

\section{Main equations of maximal slicings}
\label{sec:3}
According to the previous comment, we need to introduce two fields, $X$ and 
$Y$, such that the commutation relation $[X, Y]=0$ be satisfied. This condition
assures the existence of two coordinate parameters, namely 
$\tilde{t}=\tilde{t}(t,r)$ and $\tilde{r}=\tilde{r}(t,r)$, such that
\be
    X = \der{}{\tilde{t}},\; Y = \der{}{\tilde{r}}.
\label{e:def_X_Y}
\ee
We decompose these fields
\bea
    Y &=& \lambda \bar{Y}, \hspace{0.5cm} \bar{Y}^2=1, \\
    X &=& a \bar{Y} + b \bar{Y}^{\bot}, \hspace{0.5cm}
\bar{Y}\cdot\bar{Y}^{\bot}=0,
\eea
with $b\neq0$ and $\lambda>0$, and fixing equal to one the coefficient of
$\displaystyle \der{}{t}$ in the decomposition of $\bar{Y}^{\bot}$ in terms of
$\displaystyle \der{}{t}$ and $\displaystyle \der{}{r}$. The condition 
$\bar{Y}^2=1$ is equivalent to
\be
    \bar{Y} = f \der{}{t} + P \der{}{r}, \quad \;P = B^{-1}\left( - f C
+ \epsilon \sqrt{f^2 l^2 + B} \right),
\label{e:def_P}
\ee
being $f$ an unknown function to be determined, $\epsilon=\pm1$ and 
$l^2=-AB+C^2>0$. Moreover, $\bar{Y}\cdot\bar{Y}^{\bot}=0$ leads to
\be
    \bar{Y}^{\bot} = \der{}{t} + Q \der{}{r},\quad Q = B^{-1}\left( - C
+ \frac{\epsilon f l^2}{\sqrt{f^2 l^2 + B}} \right).
\label{e:def_Q}
\ee
Consequently, the resulting fields are
\bea
    &&X = (af+b)\der{}{t} + (aP+bQ)\der{}{r}, \label{e:dec_X} \\
    &&Y = \lambda\left(f\der{}{t} + P\der{}{r}\right)
=\lambda\left(f\alpha\,n + \frac{\epsilon}{B}\sqrt{f^2l^2+B}\,\der{}{r}\right),
\label{e:dec_Y}
\eea
where we have taken into account Eqs.~(\ref{e:def_n}) and (\ref{e:def_P}). 
Using this decomposition, we will obtain an equation, involving only the 
unknown $f$, that gives a maximal family of hypersurfaces  
$\tilde{t} = \mathrm{constant}$. Notice that Eq.~(\ref{e:dec_Y}) makes clear 
the meaning of $\epsilon$: relative to the Eulerian observer $n$, the radial 
component of $Y$ is outgoing or ingoing according to $\epsilon=1$ or 
$\epsilon=-1$.

The Jacobian matrix of the above change of coordinates can be written, in terms
of the previous decompositions, as
\bea
    \left(\begin{tabular}{cc}
                $\dot{\tilde{t}}$ & $\tilde{t}'$ \\
                $\dot{\tilde{r}}$ & $\tilde{r}'$
          \end{tabular}\right) = \left(\begin{tabular}{cc}
                                    $\displaystyle{\der{t}{\tilde{t}}}$ & 
$\displaystyle{\der{t}{\tilde{r}}}$ \vspace{0.1cm} \\
                                    $\displaystyle{\der{r}{\tilde{t}}}$ & 
$\displaystyle{\der{r}{\tilde{r}}}$
                                       \end{tabular}\right)^{-1} = \left(
\begin{tabular}{cc}
$a f + b$ & $\lambda f$ \\
$a P + b\,Q$ & $\lambda P$
\end{tabular}\right)^{-1} \nonumber \\
 = (b\; p\, \lambda)^{-1} \left(\begin{tabular}{cc}
                               $\lambda P$ & $-\lambda f$ \\
                               $-(a P + b\,Q)$ & $(a f + b)$
                               \end{tabular}\right), \;\;
\label{e:jacobian}
\eea
where $\displaystyle p = P-fQ = \frac{\epsilon}{\sqrt{f^2l^2+B}} \neq 0$. In 
order to ensure Eq.~(\ref{e:def_X_Y}), we must require that $[X,Y]=0$, and this
condition is expressed according to the following result.

{\it Lemma 1: For the fields $X,Y$ given by Eqs.~(\ref{e:dec_X}) 
and (\ref{e:dec_Y}), the condition $[X,Y]=0$ is equivalent to
\bea
     \left[\frac{P}{b\,p}\right]' = \left[\frac{-f}{b\,p}\right]
\mbox{\huge $^{^.}$}, \label{e:int1}\\
     \left[-\frac{a P + b\,Q}{b\,p\,\lambda}\right]' =
\left[\frac{a f + b}{b\,p\,\lambda}\right]\mbox{\huge $^{^.}$} .
\label{e:int2}
\eea
}

To avoid confusion, we denote with $\widetilde{K}$ the trace of the extrinsic 
curvature of the new hypersurfaces $\tilde{t}=\mathrm{constant}$. From 
Eq.~(\ref{e:K}), the maximal slicing condition for the new hypersurfaces,
$\widetilde{K} = 0$, is equivalent to
\be
    X (Y^2) + 2 \, Y^2 \, \frac{X(D)}{D} - 2 Y (X \cdot Y)
+ \left[\frac{Y(Y^2)}{Y^2} - 2 \, \frac {Y(D)}{D}\right] X \cdot Y= 0.
\label{e:pre-max_sli} \ee

Taking into account Eq.~(\ref{e:pre-max_sli}) and the commutation relation, we 
have 3 equations for 4 unknown functions, $a,b,f,\lambda$. But, due to $D$ 
transforms as a scalar under the considered coordinate transformation, we can 
add without loss of generality, the coordinate condition
\be
    \tilde{r}^2\,Y^2 = D,
\label{e:def_conf_flat}
\ee
saying that the metric on the hypersurfaces $\tilde{t} = \mathrm{constant}$ is 
written in isotropic conformally flat form. Notice that 
Eq.~(\ref{e:def_conf_flat}) guarantees that the congruence of coordinate lines 
associated with the field $\displaystyle X = \der{}{\tilde{t}}$ has minimal 
distortion.\cite{SmarrY78b,Gourg07a}

From Eq.~(\ref{e:pre-max_sli}), taking into account the conformally flat 
coordinate condition (\ref{e:def_conf_flat}), we obtain the following result.

{\it Lemma 2: The condition  that the hypersurfaces 
$\tilde{t} = \mathrm{constant}$ are maximal ($\widetilde{K} = 0$), is 
equivalent to
\be
    2 Y \left( X \cdot Y \right) - 3 X \left(Y^2\right) + 
\left[\frac{4}{\tilde{r}} + \frac{Y\left(Y^2\right)}{Y^2} \right] X\cdot Y = 0,
\label{e:max_sli}
\ee
where $X,Y$ and $\tilde{r}$ are defined by Eqs.~(\ref{e:dec_X}), 
(\ref{e:dec_Y}) and (\ref{e:def_conf_flat}).}

Lemma 2 expresses the maximal slicing condition in terms of the fields $X,Y$. 
These fields have to satisfy the commutation relation given by Lemma 1. In the 
next section, we rewrite both conditions in terms of the 
decompositions~(\ref{e:dec_X}) and (\ref{e:dec_Y}).

%%%%%%%%%%%%%%%%%%%%%%%%%%%%%%%%%%%%%%%%%%%%%%%%%%%%%%%

\section{Local existence of maximal slicings}
\label{sec:4}
In this section, we deduce the theorem stating the local existence of maximal 
slicings in SSST. The proof consists in reducing the required conditions into a
set of first order partial differential equations.

In terms of the decompositions (\ref{e:dec_X}) and (\ref{e:dec_Y}),
Eqs.~(\ref{e:def_conf_flat}) and (\ref{e:max_sli}) are expressed as
\bea
    &&\lambda = \sqrt{D} / \tilde{r} \label{e:def_lambda} \\
    &&f \dot{a} + P a' - a \left( \frac{f \dot{\lambda} + P \lambda'}{\lambda}
-\frac{2}{\sqrt{D}} \right) = 3 b \;\frac{\dot{\lambda} + Q \lambda'}{\lambda}.
\;\;\;\;\;\;\;
\label{e:lambda}
\eea
Eq.~(\ref{e:def_lambda}) can be viewed as a definition of $\tilde{r}$ in terms 
of $\lambda$, and from this equation we can derive
\be
    \frac{\lambda'}{\lambda} = \frac{D\,'}{2D} - \frac{a f + b}{b\,p \sqrt{D}},
\hspace{0.6cm} \frac{\dot{\lambda}}{\lambda} = \frac{\dot{D}}{2D}
+ \frac{a P + b\,Q}{b\,p \sqrt{D}}\, , \label{e:lambda_r_and_t}
\ee
where we have taken into account the expression of the Jacobian matrix given by
Eq.~(\ref{e:jacobian}).

At this point we have 5 equations, i.e., Eqs.~(\ref{e:int1}), (\ref{e:int2}),
(\ref{e:lambda}) and (\ref{e:lambda_r_and_t}), for 4 unknown functions, i.e.,
$(a,b,f,\lambda)$. Eqs.~(\ref{e:lambda_r_and_t}) are equivalent to give 
quantity $a$ in terms of the other unknowns,
\be
    a = b \sqrt{D} \left(\frac{\dot{\lambda}+Q\lambda'}{\lambda}
- \frac{\dot{D}+QD'}{2D} \right),
\label{e:a}
\ee
and to the following equation
\be
    f \frac{\dot{\lambda}}{\lambda} +  P \frac{\lambda'}{\lambda} =
f \frac{\dot{D}}{2D} + P \frac{D\,'}{2D} - \frac{1}{\sqrt{D}}.
\label{e:lambda_r_2}
\ee
Once $a$ has been eliminated, it can be seen that Eq.~(\ref{e:int2}) can be 
obtained from Eq.~(\ref{e:lambda_r_2}). From Eq.~(\ref{e:a}), the field $X$ 
given by Eq.~(\ref{e:dec_X}) is expressed as
\be
    X = b\,p\sqrt{D}\left(-\Lambda' \der{}{t} + \dot{\Lambda} \der{}{r}\right),
\ee
where $\displaystyle \Lambda = \log \lambda / \sqrt{D}$. So we have reduced the 
problem to solve 3 equations, i.e., Eqs.~ (\ref{e:int1}), (\ref{e:lambda}) 
(replacing $a$), and (\ref{e:lambda_r_2}), for 3 unknown functions, i.e.,
$(b,f,\lambda)$. Eq.~(\ref{e:int1}) can be rewritten as
\be
    f \frac{\dot{b}}{b} + P \frac{b\,'}{b} = P' - P \frac{p\,'}{p} + \dot{f}
- f \frac{\dot{p}}{p} \,.
\label{e:int1_2}
\ee
As it will be shown below in Eq.~(\ref{e:int1_3}), $f$ can be obtained from an 
equation which does not contain other variables. Then, Eq.~(\ref{e:int1_2}) is 
a first order partial differential equation (PDE) for $b$.

Eliminating $a$ and its derivatives with Eq.~(\ref{e:a}), and making use of
Eq.~(\ref{e:int1_2}) to eliminate $b$, Eq.~(\ref{e:lambda}) is reduced to a 
second order PDE involving $\lambda$ and $f$. Finally, making use of 
Eq.~(\ref{e:lambda_r_2}) for, firstly, reducing the order of the equation and, 
secondly, for eliminating $\lambda$, Eq.~(\ref{e:lambda}) is equivalent to the
following one:
\be
    \frac{\dot{p}}{p} - \frac{\dot{D}}{D} - Q'
+ Q \left[ \frac{p\,'}{p} - \frac{D'}{D} \right] = 0.
\label{e:int1_3}
\ee
Notice that previous equation involves only $f$ when $p$ and $Q$ are written 
explicitly in terms of $f$ from their definitions.

We have found a decoupled system of quasi-linear PDEs (\ref{e:lambda_r_2}), 
(\ref{e:int1_2}) and (\ref{e:int1_3}) according to the following steps: First, 
Eq.~(\ref{e:int1_3}) can be solved for $f$. Second, Eq.~(\ref{e:lambda_r_2}) 
can be solved for $\lambda$. Third, Eq.~(\ref{e:int1_2}) can be solved for $b$.
Finally, $a$ can be obtained from Eq.~(\ref{e:a}). Assuming that all the metric
components $A, B, C, D$ are continuously differentiable functions, the initial 
value problem with respect to this set of equations has always local (both in 
space and time) solution.\cite{pde} We have reached the main objective of this
section, that is to prove the following result.

{\it Theorem: Any spherically symmetric spacetime can be locally 
sliced by a family of maximal spacelike hypersurfaces.}

Let us make a few comments about first order partial differential equations. If
the functions appearing in the equation are continuously differentiable and we 
are looking for smooth solutions, given continuously differentiable initial 
data, it always exists a solution of the equation in a domain containing the 
given initial data. The domain of the solution can be the domain of initial 
parameters or a smaller one; so that is why, in principle, one can only 
guarantee local existence of solutions. Moreover, in this kind of equations one
can use the so-called ``characteristic method'' in order to obtain, analytical 
or numerically, solutions of the equation. This method is based on two facts: 
i) given initial data at a point, it always exists a characteristic curve 
containing the given initial data; ii) a solution of the equation can be 
constructed by the union of a set of characteristic curves. So, a solution of 
the equation can be obtained by solving a set of characteristic curves around a
given point.

Let us also point out that in the resolution of the characteristic curves some 
functional relations involving the dependent and independent variables have to 
be constant; the cardinal of the set of independent functional relations 
depends on the number of independent variables of the equation, and the general
solution of the first order partial differential equation can be written as an 
implicit function of those ones.

Once the local existence of maximal slicings has been proved, we are going (see
next Section) to design a practical procedure to build up these slicings.

%%%%%%%%%%%%%%%%%%%%%%%%%%%%%%%%%%%%%%%%%%%%%%%%%%%%%%%

\section{A procedure to construct maximal slicings}
\label{sec:5}
The proof of the above theorem was based on the explicit finding of the minimum
set of equations to be solved in order to obtain maximal slicings in SSSTs. Let
us analyze a constructive algorithm that complements the theorem. We consider 
two different cases, $f=0$ and $f\neq0$.

%%%%%%%%%%%%%%%%%%%%

\subsection{First case $f=0$}

If $f=0$, Eq.~(\ref{e:int1_3}) is reduced to $K=0$, the maximal slicing 
condition provided by Eq.~(\ref{e:K}). In this case, from 
Eqs.~(\ref{e:int1_2}), (\ref{e:lambda_r_2}) and (\ref{e:a}), it turns out that 
$b(t)\neq 0$ arbitrary, $\lambda = Z(t) \, \sqrt{D} \, 
\exp \left ( - \epsilon \int \limits_{R}^r\sqrt{B/D}dr\right)$, with $Z(t)>0$ 
arbitrary and $R$ constant, and
\be
	a = b \sqrt{D}\left[ \dot{Z}/Z + \epsilon\, C / \sqrt{BD} - \epsilon
\int \limits_{R}^r \left( \sqrt{B/D} \right) \hspace{-1mm}\mbox{\huge$^.$}\, dr\right].
\nonumber
\ee
The fields $X$ and $Y$ can be integrated according to:

{\it Proposition 1: If the hypersurfaces $t=\mathrm{constant}$ in 
the spacetime defined by the metric~(\ref{e:sph_sym}) are maximal, the 
following change of coordinates
\be
 \tilde{t} = \tilde{t}(t), \hspace{0.5cm} \tilde{r} = Z^{-1}\,
\exp\left(\epsilon\int\limits_R^r\sqrt{B/D} \, dr\right),
\ee
with $\dot{\tilde{t}}=b^{-1}$, $Z(t)>0$ and $R$ constant, allows one to write 
the metric in isotropic form
\bea
	ds^2 &=& b^2 \left[A + D\left(\frac{\dot{\tilde{r}}}{\tilde{r}}\right)^2 -
2\epsilon \, C \sqrt{\frac{D}{B}} \,\, \frac{\dot{\tilde{r}}}{\tilde{r}}\right] 
d\tilde{t}^{\,2} + 2\, \frac{\epsilon \, b}{\tilde{r}}\sqrt{\frac{D}{B}}
\left(C-\epsilon\sqrt{BD}\, \, 
\frac{\dot{\tilde{r}}}{\tilde{r}}\right) d\tilde{t} \, d\tilde{r} \nonumber \\
	&&+ \frac{D}{\tilde{r}^2}\left(d\tilde{r}^2+\tilde{r}^2d\Omega^2\right).
\label{e:metric_f=0}
\eea
}
This metric form is adapted to the same initial maximal slices, that are now 
labeled as $\tilde{t}=\mathrm{constant}$. It remains invariant under the 
inversion $\displaystyle\tilde{r} \rightarrow T(\tilde{t})/\tilde{r}$ when, in 
addition, the sign of the radial component of the field $Y$ is reversed by the 
change $\epsilon\rightarrow-\epsilon$.

%%%%%%%%%%%%%%%%%%%%%%%%%%%%%

\subsection{Second case $f\neq 0$}
When $f\neq0$, the explicit expressions of $P/f$ and $Q$ in terms of $f$, allow
us to define the variable
\be
    F = \frac{\epsilon f}{\sqrt{f^2l^2+B}} \quad \Leftrightarrow \quad
f = \epsilon F\sqrt{\frac{B}{1-l^2F^2}}.
\label{e:def_F}
\ee
Then, using the definition of $P$ and $Q$ given by Eqs.~(\ref{e:def_P}) and
(\ref{e:def_Q}) respectively, we have:
\be
    \frac{P}{f}=\frac{1-FC}{FB}, \qquad Q=\frac{-C+Fl^2}{B}, \qquad 
p=\epsilon\sqrt{\frac{1-l^2F^2}{B}}.
\label{e:P_Q_p}
\ee
Eqs.~(\ref{e:int1_2}), (\ref{e:lambda_r_2}) and (\ref{e:int1_3}) are rewritten,
respectively, as
\bea
    &&\dot{b} + \frac{1-FC}{FB}b' = b \left\{ \frac{\dot{B}-C'}{B}
+ \frac{1}{1-l^2F^2} \left[ - (1+l^2F^2) \, \frac{\dot{F}}{F}
+ \frac{l^2F(2-FC)-C}{B} \, \frac{F'}{F}\nonumber \right.\right.\\
    &&\left.\left.\hspace{2.8cm} + F^2(l^2)\dot{} + \frac{(1-FC)F(l^2)'}{B} 
\right] \right\}, \label{e:system1} \\[4mm]
    &&\dot{\lambda} + \frac{1-FC}{FB}\lambda' = \lambda\left[\frac{\dot{D}}{2D}
+ \frac{1-FC}{FB} \frac{D'}{2D} -
\frac{\epsilon\sqrt{1-l^2F^2}}{F\sqrt{BD}} \right],
\label{e:system2}\\[4mm]
    &&\dot{F} + \frac{1-FC}{FB}F' =
- \frac{FB(l^2)\dot{}+(2-l^2F^2-FC)(l^2)'}{2l^2B} \nonumber \\
    &&\hspace{2.8cm} - L\left[\frac{\dot{B}}{2}+\frac{B\dot{D}}{D}
    +\frac{(C-Fl^2)(DB'-2BD')}{2BD} - C'\right], \label{e:system3}
\eea
where $L=(1-l^2F^2)/(l^2FB)$. The solution can be obtained numerically solving 
the quasilinear Eqs.~(\ref{e:system3}), (\ref{e:system2}) and (\ref{e:system1})
for, respectively, $F$, $\lambda$ and $b$, given initial data for $t=t_0$. 
These equations are of hyperbolic type, analogous to the well-known advective 
equation in classical fluid dynamics. The three equations have a common 
``advective velocity": $\displaystyle V = (1-FC)/(FB)$. This velocity 
can be interpreted as the radial component of the light velocity in the 
modified metric $\displaystyle d\tilde{s}^2 = ds^2 - \frac{1}{f^2}dt^2$.

On the other hand, the coordinate 1-forms $d\tilde{t}$ and $d\tilde{r}$ are 
easily written from Eq.~(\ref{e:jacobian}), taking into account 
Eqs.~(\ref{e:P_Q_p}). Then, we arrive to this result:

{\it Proposition 2:  For any given SSST a maximal slicing can be 
built up according to the following steps:
\begin{itemize}
\item [(i)] To start with the metric form~(\ref{e:sph_sym}), with
$l^2 \equiv C^2-AB>0$, $B>0$ and $D>0$, to solve Eq.~(\ref{e:system3}) for $F$,
and then, to solve Eq.~(\ref{e:system1}) for $b$.
\item [(ii)] To define
\be
    \Gamma \equiv b^{-1}(1-l^2F^2)^{-1}\left[(1-FC)dt - FB dr\right]
\label{e:def_Gamma}
\ee
which is necessarily a closed 1-form.
\item [(iii)] To find a potential $\tilde{t}(t,r)$ of $\Gamma$, 
$\Gamma = d\tilde{t}$. Then, the level surfaces of $\tilde{t}$ are spacelike 
and maximal.
\end{itemize}
}

Analogously, by applying Eq. (\ref{e:def_lambda}), $\tilde{r}$ can be 
integrated once $F$ and $\lambda$ are solved. The change of coordinates
$\left\{ \tilde{t} = \tilde{t}(t,r), \tilde{r} = \tilde{r}(t,r), \theta, 
\varphi \right\}$ is completely defined from the fields $X,Y$ given by 
Eqs.~(\ref{e:dec_X}) and (\ref{e:dec_Y}). Then, the metric form can be deduced 
in the new set of coordinates. The new slices are maximal and Proposition 1 can
again be applied.

%%%%%%%%%%%%%%%%%%%%%%%%%%%%%%%%%%%%%%%%%%%%%%%%%%%%%%%

\section{Maximal slicings in flat spacetime}
\label{sec:6}
In order to illustrate how the procedure designed in previous section works, 
let us consider the simple case of the Minkowski spacetime starting with 
inertial spherical coordinates, $ds^2 = - dt^2 + dr^2 + r^2 d\Omega^2$. 
According to Proposition 1, $\tilde{t}=\tilde{t}(t)$, with 
$\dot{\tilde{t}}=b^{-1}$, $\tilde{r}=r^{\epsilon}\,Z^{-1}$, and the resulting 
metric is
\bea
	ds^2 = \left[-b^2 +
\left(\tilde{r}^{\epsilon}\,Z^{\epsilon-1}\frac{dZ}{d\tilde{t}}\right)^2\right]
d\tilde{t}^{\,2} + 2\left(\tilde{r}\,Z\right)^{2\epsilon-1}\frac{dZ}{d\tilde{t}} 
d\tilde{t}\,d\tilde{r} 
+\left(\tilde{r}^{\epsilon-1}\,Z^{\epsilon}\right)^2
\left(d\tilde{r}^2+\tilde{r}^2d\Omega^2\right),
\eea
$b=b(\tilde{t})\neq0,Z=Z(\tilde{t})>0$ being arbitrary functions. So, we have 
the same slicing that the initial inertial one (this foliation is totally 
geodesic, i.e., its extrinsic curvature tensor vanishes). However, other 
maximal slicings (non-inertial, in general) can be obtained as a consequence of
Proposition 2. According to it, Eq.~(\ref{e:system3}) reads
\be
    \dot{F} + F^{-1}\,F' = -2r^{-1}(1-F^2).
\ee
Following the last comments of Sec.~\ref{sec:4}, the general solution of this 
equation can be given in terms of an implicit function,
\be \displaystyle \Phi_1 \left( m(F) + t \;r^{-1}
\left(F^{-2}-1\right)^{1/4}, r^{-1}\left(F^{-2}-1\right)^{1/4}
\right) = 0, \nonumber
\ee
with $0<F^2<1$, 
$m(F)=4\sqrt{|F|}\;\mathrm{_2F_1} \left(1/4,\;3/4;\;5/4;\;F^2\right)$, 
$\mathrm{_2F_1}$ being the classical standard hypergeometric series 
\cite{hypergeo} and $\Phi_1$ being an arbitrary function. Once $F$ is known, 
the general solution of Eqs.~(\ref{e:system1}) and (\ref{e:system2}) are, 
respectively, $\Phi_2(I_1,I_2)$ and $\Phi_3(I_1,I_3)$, where $\Phi_2$ and 
$\Phi_3$ are arbitrary functions, and
\bea
	&&I_1 = r-\int\limits_0^tF^{-1}dt, \\
  &&I_2 = b^{-1} \exp \left[ \int \limits_0^t
\left( -\frac{(1+F^2)\dot{F}}{F}+2F' \right) \frac{dt}{1-F^2} \right], \\
  &&I_3 = \lambda^{-1}\exp\left(r^{-1}\int\limits_0^t
F^{-1}(1-\epsilon\sqrt{1-F^2})dt\right).
\eea
This study can be extended to the Schwarzschild spacetime using different 
metric forms, specially those that are regular at the horizon, like the 
Painlev\'e-Gullstrand form and other similar ones.\cite{PaFont} Maximal 
slincings can also be constructed in other spacetimes 
(Reissner-Nordtr$\ddot{\rm o}$m, Vaidya, Lema{\^{\i}}tre-Tolman-Bondi, etc.) 
and the results compared with previous works.\cite{Estabrook-WCWST73,Beig-Mu98,Reinhart73,Malec94,PetriST85}

%%%%%%%%%%%%%%%%%%%%%%%%%%%%%%%%%%%%%%%%%%%%%%%%%%%%%%%

\section{Maximal slicings in closed Friedmann spacetime}
\label{sec:7}
The pioneering analysis by Oppenheimer and Snyder\cite{OppSny} on the dynamics
of the collapse of a homogeneous sphere of pressure-zero matter allowed them to
obtain an analytical solution of Einstein equations for a dynamical spacetime. 
Its metric, a special case of Lema{\^{\i}}tre-Tolman-Bondi metric, has a simple
form (see below), being the Friedmann metric in the interior. We are going to 
apply the procedure designed in Sec.~\ref{sec:5} to build maximal slicings in a
closed Friedmann spacetime. We start with the following metric form:
\be
	ds^2 = R^2 \left(- d\eta^2 + d\chi^2 + \sin^2\chi d\Omega^2\right),
\label{e:dust_metric}
\ee
where $R = R_0 \, (1-\cos\eta)$, $R_0$ being a positive constant, and 
$\eta\in(-\pi,0),\, \chi\in(0,\pi)$, being the domain of the parameters ($\eta$ 
and $\chi$ will play the role of $t$ and $r$ in Eq.~(\ref{e:sph_sym}) 
respectively). The hypersurfaces $\eta=$ constant are not maximal. We are going
to solve numerically Eqs.~(\ref{e:int1_3}), (\ref{e:int1_2}) and 
(\ref{e:lambda_r_2}). Let us define $\tilde{f} := R f$.

\begin{figure}
\begin{center}
\includegraphics[width=8cm,height=6cm]{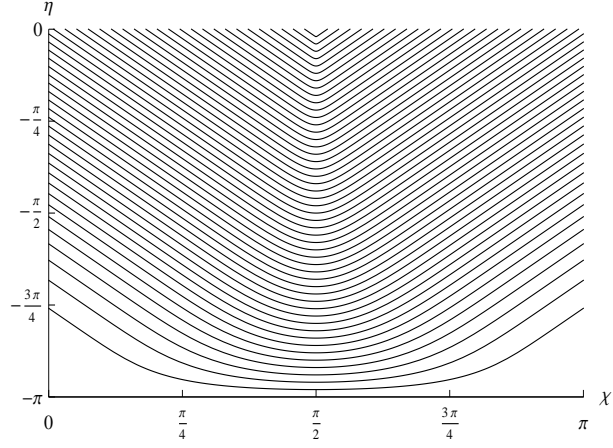}
\caption{Projection onto the $\eta-\chi$ plane of characteristic curves of 
Eqs.~(\ref{e:charac_f}), with $\tilde{f}(\chi=\pi/2)=0$, 
$\eta(\chi=\pi/2)=-k\,\pi$, $0 < k \leq 1$, initial data.}
\label{fig:charac_eta_chi}
\end{center}
\end{figure}

\begin{figure}
\begin{center}
\includegraphics[width=9cm,height=6cm]{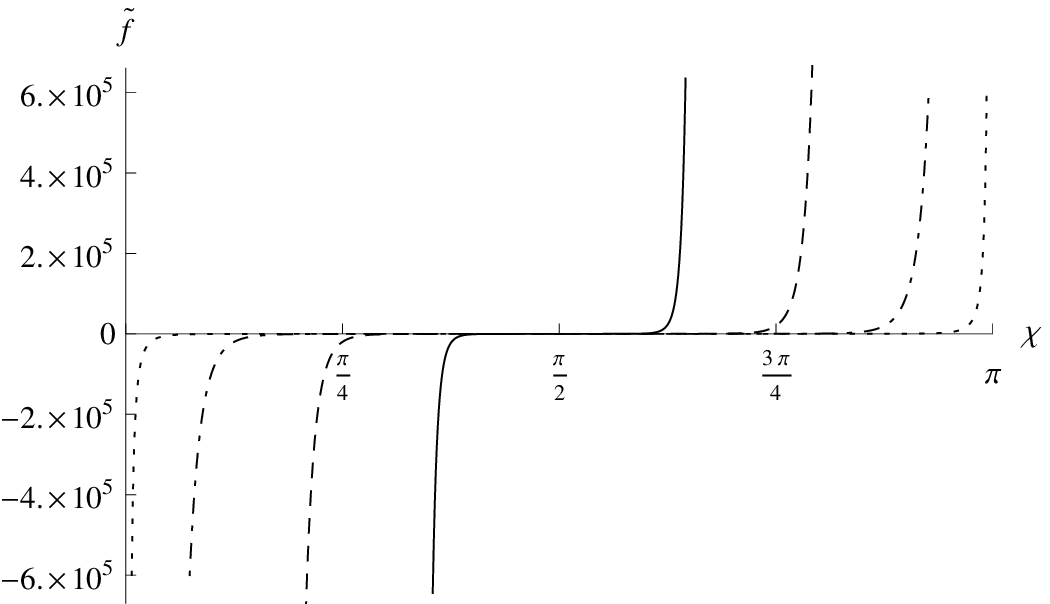}
\caption{Projection onto the $\tilde{f}-\chi$ plane of characteristic curves of 
Eqs.~(\ref{e:charac_f}), with $\tilde{f}(\chi=\pi/2)=0$, 
$\eta(\chi=\pi/2)=-k\,\pi$, $k$ = 1/5 (solid line), 2/5 (dashed line), 3/5 
(dash-dotted line), 4/5 (dotted line) initial data.}
\label{fig:charac_tildef_chi}
\end{center}
\end{figure}

Eq.~(\ref{e:int1_3}) reads
\be
	\frac{\tilde{f}}{1+\tilde{f}^2} \frac{\partial\tilde{f}}{\partial\eta} + 
\frac{\epsilon}{\sqrt{1+\tilde{f}^2}} \frac{\partial\tilde{f}}{\partial\chi} + 
6 \cot\frac{\eta}{2} + \frac{2\epsilon\tilde{f}\cot\chi}{\sqrt{1+\tilde{f}^2}}
= 0.
\ee
Without loss of generality, it can be considered $\epsilon=1$, since for 
$\epsilon=-1$ the solution will be $-\tilde{f}$. Previous equation is then 
equivalent to
\be
	\frac{\tilde{f}}{\sqrt{1+\tilde{f}^2}} \frac{\partial\tilde{f}}{\partial\eta}+
\frac{\partial\tilde{f}}{\partial\chi} = - 6 \sqrt{1+\tilde{f}^2} \, 
\cot\frac{\eta}{2} - 2 \tilde{f} \, \cot\chi.
\label{e:dust_tildef}
\ee
We use the characteristic method for solving the above equation 
(\ref{e:dust_tildef}). Let $(\eta(\mu),\chi(\mu),\tilde{f}(\mu))$ be a 
characteristic curve depending on a parameter $\mu$ such that 
$\displaystyle \frac{d\eta}{d\mu}=\frac{\tilde{f}}{\sqrt{1+\tilde{f}^2}}$, 
\, $\displaystyle \frac{d\chi}{d\mu}=1$, and $\displaystyle \frac{d\tilde{f}}{d\mu}
= - 6 \sqrt{1+\tilde{f}^2} \, \cot\frac{\eta}{2} - 2 \tilde{f} \, \cot\chi$. 
Substituting $d\chi=d\mu$, and considering then $(\eta(\chi),\tilde{f}(\chi))$,
the characteristic system associated to Eq.~(\ref{e:dust_tildef}) is written as
\be
	d\eta / d\chi = \tilde{f} / \sqrt{1+\tilde{f}^2}, \,\,
d\tilde{f} / d\chi = - 6\sqrt{1+\tilde{f}^2}\,\cot\frac{\eta}{2}
- 2\tilde{f}\,\cot\chi.
\label{e:charac_f}
\ee
Notice that, although $\eta=-\pi$ is out of the range of the parameter, 
$\tilde{f}=0, \, \eta=-\pi$ is a solution of the previous system. Notice also 
that solutions of Eqs.~(\ref{e:charac_f}) satisfy $\displaystyle \tilde{f} 
\left( \frac{\pi}{2} - x \right) = -\tilde{f} \left( \frac{\pi}{2} + x \right)$
and $\displaystyle \eta \left( \frac{\pi}{2} - x \right) = \eta \left( 
\frac{\pi}{2} + x \right)$. Eqs.~(\ref{e:charac_f}) can be solved numerically, 
imposing
\be
	\tilde{f}(\chi=\pi/2)=0,\, \eta(\chi=\pi/2)=-k\,\pi,\, 0 < k \leq 1,
\ee
as initial data. In Fig.~\ref{fig:charac_eta_chi} we plot the projection onto
the $\eta-\chi$ plane of the characteristic curves, for the values $k=-l/50$, 
$l=1,\ldots,50$. In Fig.~\ref{fig:charac_tildef_chi} we plot the projection 
onto the $\tilde{f}-\chi$ plane of the characteristic curves, for the values 
$k = l/5$, $l=1,\ldots,4$. The union of the characteristic curves, for all the 
possible values of $k, 0 < k \leq 1$, covers the whole domain of parameters for
$\eta$ and $\chi$. This union constitutes a global solution of 
Eq.~(\ref{e:dust_tildef}) for $\tilde{f}$.

\begin{figure}
\begin{center}
\includegraphics[width=8cm,height=6cm]{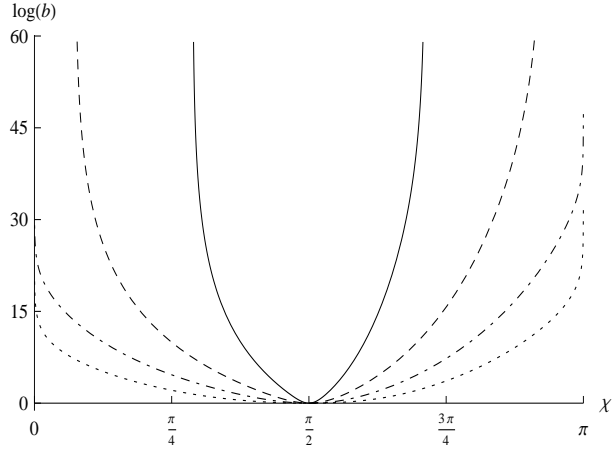}
\caption{Projection onto the $\log b-\chi$ plane of characteristic curves of 
Eqs.~(\ref{e:charac_logb}), with $\log b\,(\chi=\pi/2)=0$, previous values for 
$\tilde{f}$ and $\eta$, $k$ = 1/5 (solid line), 2/5 (dashed line), 3/5 
(dash-dotted line), 4/5 (dotted line), initial data.}
\label{fig:charac_logb_chi}
\end{center}
\end{figure}

Once $\tilde{f}$ (and therefore $f$) is solved, we focus on 
Eq.~(\ref{e:int1_2}) for $b$, which reads (after multiplying by $\epsilon R$)

\bea
	\epsilon \tilde{f} \, \frac{\partial\log b}{\partial\eta} + 
\sqrt{1+\tilde{f}^2} \frac{\partial\log b}{\partial\chi}
	= \frac{\partial\tilde{f}}{\partial\eta} \left[ 
\frac{\tilde{f}}{\sqrt{1+\tilde{f}^2}} + 
\epsilon \left( 1+\frac{\tilde{f}^2}{1+\tilde{f}^2} \right) \right] + 
\frac{\tilde{f}}{\sqrt{1+\tilde{f}^2}} \frac{\partial\tilde{f}}{\partial\chi}. 
\eea
From Eq.~(\ref{e:dust_tildef}), the combinations $\epsilon\tilde{f}$ and 
$\tilde{f}^2$ will remain the same for both values of $\epsilon$. Without loss 
of generality, we take $\epsilon=1$ and the previous solution for $\tilde{f}$. 
Due to numerical reasons, we substitute (using Eq.~(\ref{e:dust_tildef}))
$\displaystyle \frac{\partial\tilde{f}}{\partial\chi}$ in terms of 
$\displaystyle \frac{\partial\tilde{f}}{\partial\eta}$, leading to
\bea
	\frac{\tilde{f}}{\sqrt{1+\tilde{f}^2}} \frac{\partial\log b}{\partial\eta} +
\frac{\partial\log b}{\partial\chi} &=& 
\frac{\partial\tilde{f}/\partial\eta}{\sqrt{1+\tilde{f}^2}} \left(\frac{\tilde{f}}{\sqrt{1+\tilde{f}^2}} + 1\right)\nonumber \\
        && - \frac{2\tilde{f}^2}{1+\tilde{f}^2} \cot\chi  - \frac{6\tilde{f}}{\sqrt{1+\tilde{f}^2}} \, \cot\frac{\eta}{2}.
\label{e:dust_logb}
\eea
We use again the characteristic method for solving the above equation 
(\ref{e:dust_logb}). Let us consider a characteristic curve 
$(\eta(\chi),\log b(\chi))$, such that
\bea
	d\eta / d\chi &=& \tilde{f} / \sqrt{1+\tilde{f}^2}, \nonumber \\
\frac{d\log b}{d\chi} &=& 
\frac{\partial\tilde{f}/\partial\eta}{\sqrt{1+\tilde{f}^2}} 
\left(\frac{\tilde{f}}{\sqrt{1+\tilde{f}^2}} + 1\right) - 
\frac{2\tilde{f}^2}{1+\tilde{f}^2} \cot\chi 
- \frac{6\tilde{f}}{\sqrt{1+\tilde{f}^2}} \, \cot\frac{\eta}{2}.\,\,
\label{e:charac_logb}
\eea
Notice that the projection onto the $\eta-\chi$ plane will be exactly the same 
as the one for the characteristic curves associated to $\tilde{f}$. In 
practice, the characteristic curves for $\log b$ can be calculated 
simultaneously to the ones for $\tilde{f}$. Notice also that, although 
$\eta=-\pi$ is out of the range of the parameter, $b = 1, \, \eta=-\pi$ (with 
$\tilde{f}=0$) is a solution of the previous system. Eqs.~(\ref{e:charac_logb})
can be solved numerically, imposing
\be
	\log b \, (\chi=\pi/2) = 0,
\ee
and the previous values for $\tilde{f}$ and $\eta$ as initial data. In 
Fig.~\ref{fig:charac_logb_chi} we plot the projection onto the $\log b-\chi$ 
plane of the characteristic curves, for the values $k = l/5$, $l=1,\ldots,4$. 
The union of the characteristic curves, for all the possible values of 
$k, 0 < k \leq 1$, covers the whole domain of parameters for $\eta$ and $\chi$.
This union constitutes a global solution of Eq.~(\ref{e:dust_logb}) for 
$\log b$. $b\neq0$ is satisfied.

\begin{figure}
\begin{center}
\subfigure[\,$\epsilon=1$.]{\includegraphics[width=7cm,height=5cm]{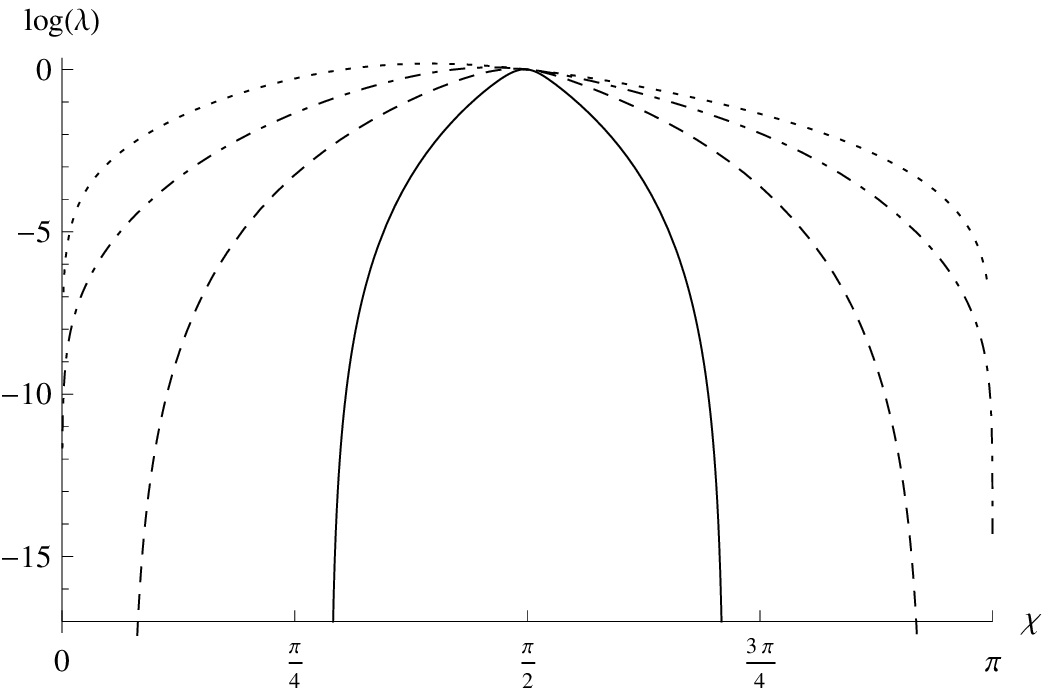}}\quad
\subfigure[\,$\epsilon=-1$.]{\includegraphics[width=7cm,height=5cm]{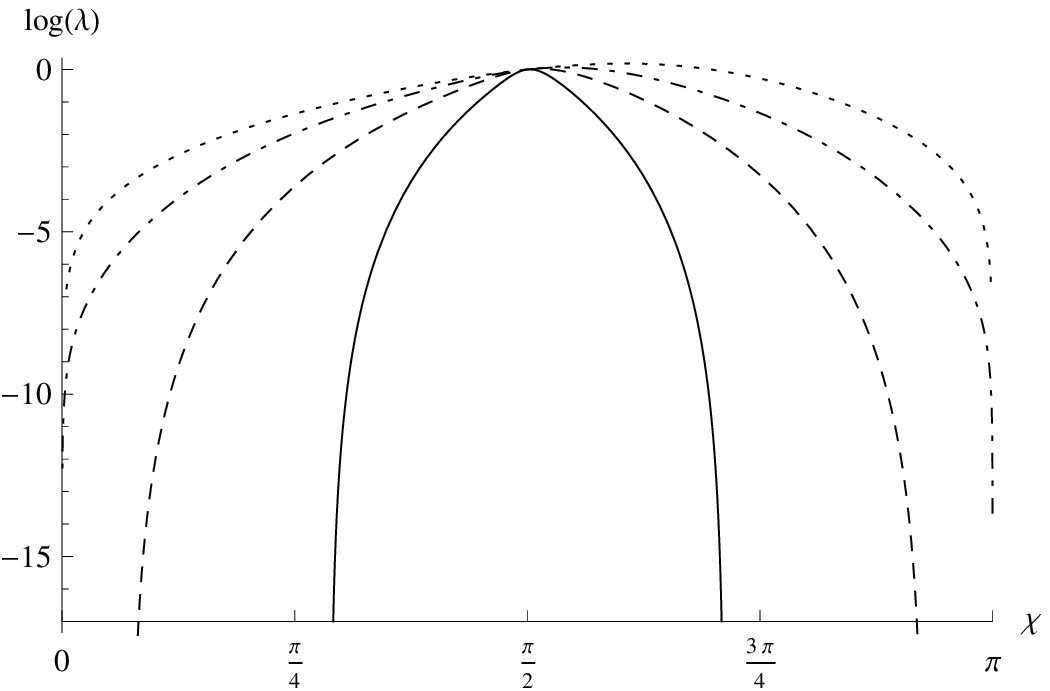}}
\caption{Projection onto the $\log\lambda-\chi$ plane of characteristic curves 
of Eqs.~(\ref{e:charac_loglambda}), with $\log\lambda\,(\chi=\pi/2)=0$, 
previous values for $\tilde{f}$ and $\eta$, $k$=1/5 (solid line), 2/5 (dashed 
line), 3/5 (dash-dotted line), 4/5 (dotted line), initial data.}
\label{fig:charac_loglambda_chi}
\end{center}
\end{figure}

We focus now on (\ref{e:lambda_r_2}) for $\lambda$, which reads (after 
multiplying by $\epsilon R/\sqrt{1+\tilde{f}^2}$\,)
\be
	\frac{\epsilon\tilde{f}}{\sqrt{1+\tilde{f}^2}}
\frac{\partial\log \lambda}{\partial\eta}
+ \frac{\partial\log \lambda}{\partial\chi}
= \frac{2\epsilon\tilde{f}}{\sqrt{1+\tilde{f}^2}} \cot\frac{\eta}{2}
+ \frac{\cos\chi - \epsilon / \sqrt{1+\tilde{f}^2}}{\sin\chi}.
\ee
The combinations $\epsilon\tilde{f}$ and $\tilde{f}^2$ will remain the same for
both values of $\epsilon$, so we can use previous solution for $\tilde{f}$ but 
we have to keep $\epsilon$ in the last term,
\be
	\frac{\tilde{f}}{\sqrt{1+\tilde{f}^2}} 
\frac{\partial\log \lambda}{\partial\eta} + 
\frac{\partial\log \lambda}{\partial\chi} =
\frac{2\tilde{f}}{\sqrt{1+\tilde{f}^2}} \cot\frac{\eta}{2} + \frac{\cos\chi 
- \epsilon / \sqrt{1+\tilde{f}^2}}{\sin\chi}.
\label{e:dust_loglambda}
\ee
Therefore, we will have two solutions $\log \lambda_{\epsilon}$ which do not 
differ only in the sign. Using again the characteristic method for solving the 
above equation (\ref{e:dust_loglambda}), we consider a characteristic curve 
$(\eta(\chi),\log \lambda(\chi))$, such that
\be
	d\eta / d\chi = \tilde{f} / \sqrt{1+\tilde{f}^2}, \,\,\,
	\frac{d\log \lambda}{d\chi} = \frac{2\tilde{f}}{\sqrt{1+\tilde{f}^2}}
\cot\frac{\eta}{2} + \frac{\cos\chi - \epsilon/\sqrt{1+\tilde{f}^2}}{\sin\chi}.
\label{e:charac_loglambda}
\ee
The projection onto the $\eta-\chi$ plane is exactly the same as the one for 
the other characteristic curves. Hence, the corresponding projections for 
$\log \lambda$ can be calculated simultaneously to them. Notice that, although 
$\eta=-\pi$ is out of the range of the parameter, 
$\lambda_{\epsilon} = \lambda_0 \, \sin\chi \, (\cot\chi/2)^{\epsilon}$, 
$\eta=-\pi$, $\lambda_0$ being a positive constant, is a solution of the 
previous system (with $\tilde{f}=0$). Eqs.~(\ref{e:charac_loglambda}) can be 
solved numerically, imposing
\be
	\log \lambda \, (\chi=\pi/2) = 0,
\ee
and the previous values for $\tilde{f}$ and $\eta$ as initial data. These 
initial data corresponds to the value $\lambda_0=1$ in the limit $\eta = -\pi$.
In Fig.~\ref{fig:charac_loglambda_chi} we plot the projection onto the 
$\log \lambda-\chi$ plane of the characteristic curves, for the values 
$k = l/5$, $l=1,\ldots,4$, and for both values of $\epsilon$. The union of the 
characteristic curves, for all the possible values of $k, 0 < k \leq 1$, covers
the whole domain of parameters for $\eta$ and $\chi$. This union constitutes a 
global solution of Eq.~(\ref{e:dust_loglambda}) for $\log\lambda$. $\lambda>0$ 
is satisfied.

Notice that the fact that the projection onto the $\eta-\chi$ plane of the 
characteristic curves for $\tilde{f}$, $\log b$ and $\log\lambda$ coincides is 
just a consequence of the fact that Eqs.~(\ref{e:system1})--(\ref{e:system3}) 
share the same advective velocity.

Once $\tilde{f}$, $b$ and $\lambda$ have been derived, we can calculate the 
unknown $a$ from Eq.~(\ref{e:a}),
\be
	a = \frac{b\,R}{1+\tilde{f}^2} \left[ \sin\chi \left( 
\partial \log \lambda / \partial \eta - 2\cot\eta/2\right) - \tilde{f} \right].
\ee

\begin{figure}
\begin{center}
\subfigure[$\,k=1/5$.]{\includegraphics[width=6.1cm,height=5.2cm]{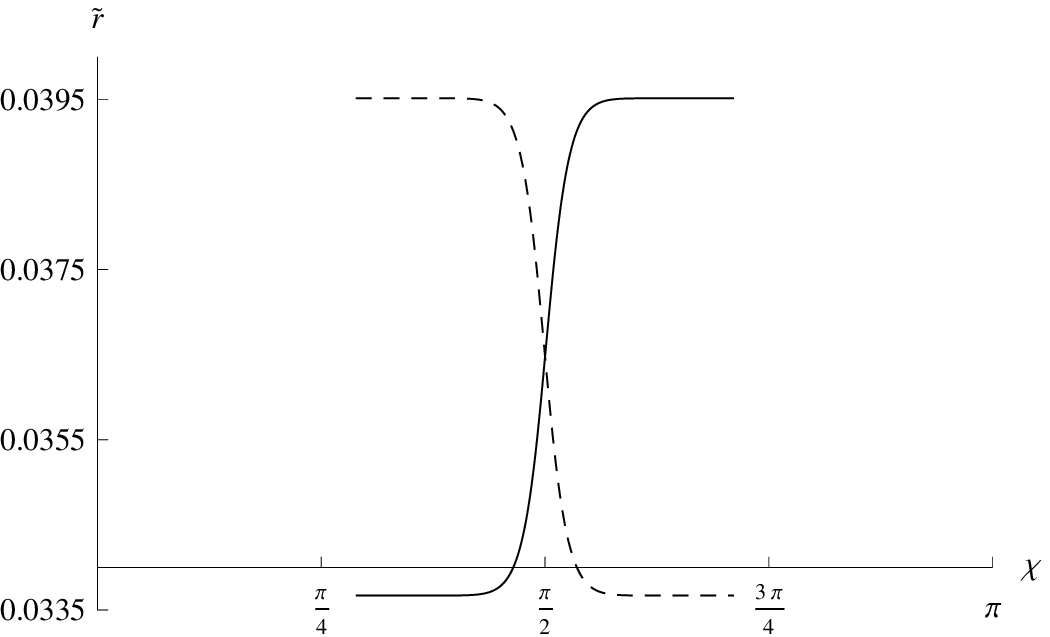}}\quad
\subfigure[$\,k=2/5$.]{\includegraphics[width=6.1cm,height=5.2cm]{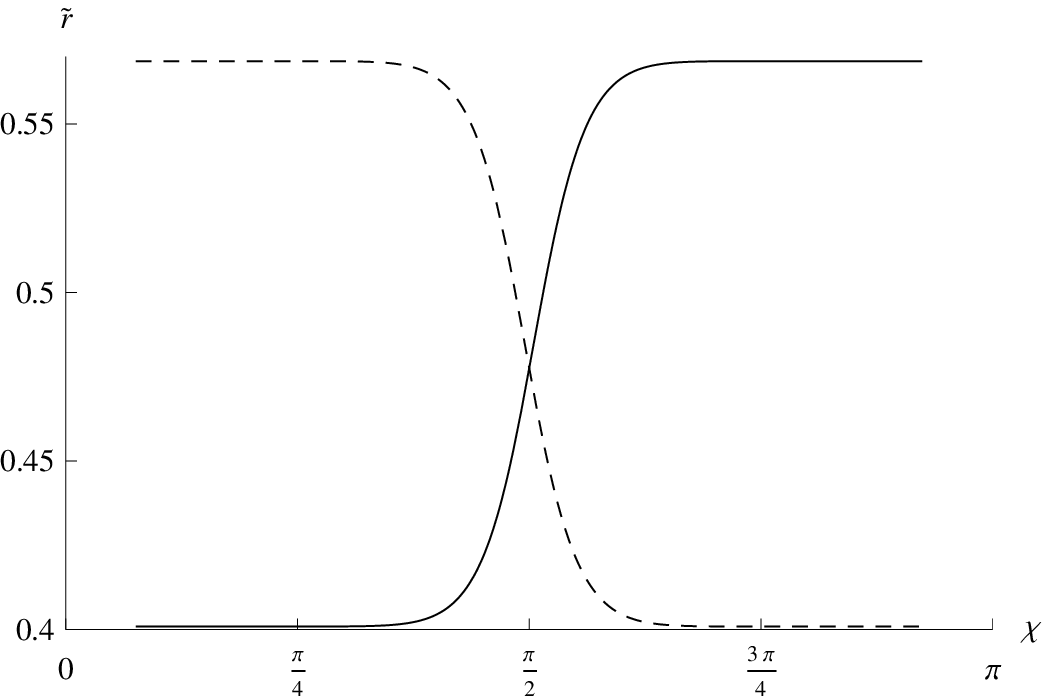}}\quad
\subfigure[$\,k=3/5$.]{\includegraphics[width=6.1cm,height=5.2cm]{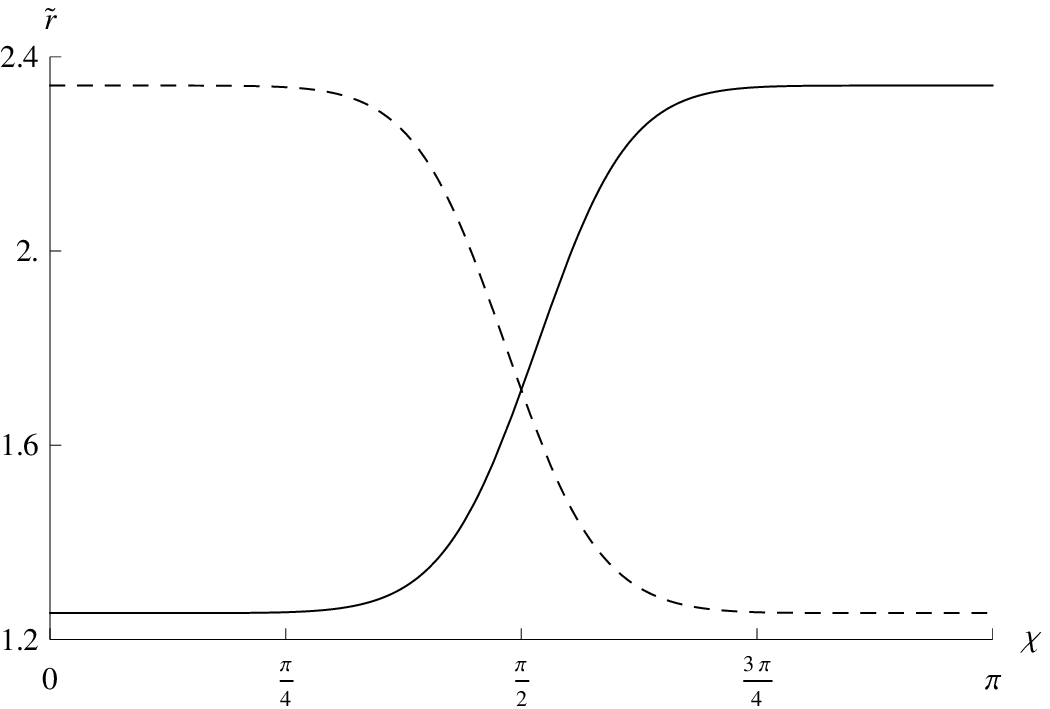}}\quad
\subfigure[$\,k=4/5$.]{\includegraphics[width=6.1cm,height=5.2cm]{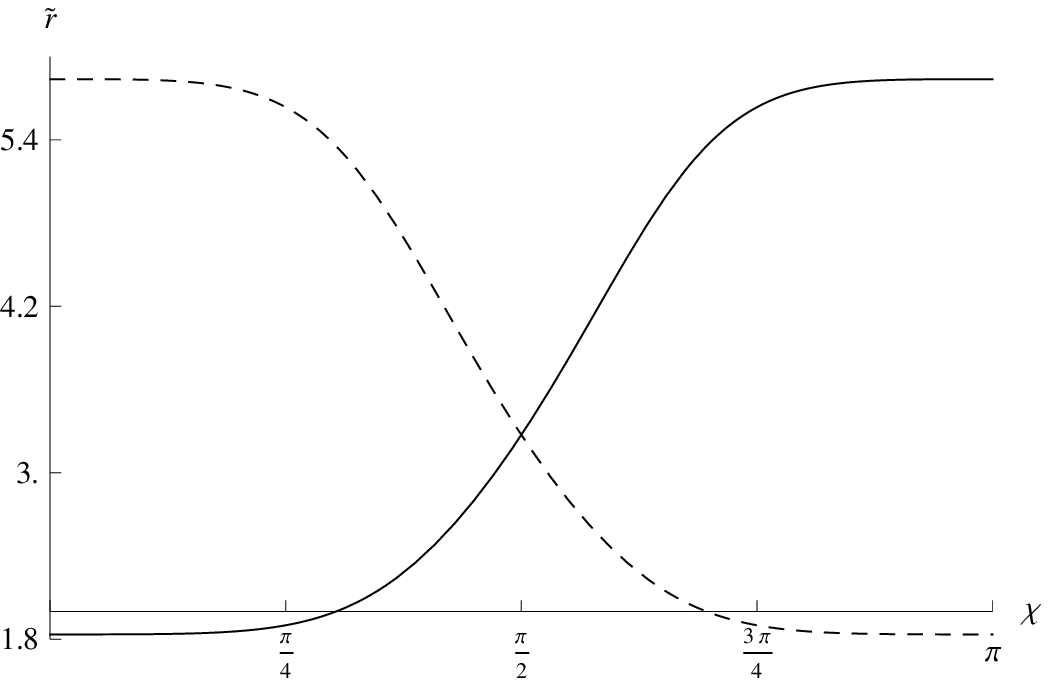}}
\caption{Projection of characteristic curves onto $\tilde{r}-\chi$ plane. 
Solid lines correspond to $\epsilon=1$, while dashed lines correspond to 
$\epsilon=-1$.}
\label{fig:rtilde_chi}
\end{center}
\end{figure}

Also, from Eq.~(\ref{e:def_lambda}), we can directly obtain
\be
	\tilde{r} = \frac{R}{\lambda}\,\sin\chi \, .
\ee
In Fig.~\ref{fig:rtilde_chi} we plot the projection onto the $\tilde{r}-\chi$
plane of the characteristic curves, for the values $k = l/5$, $l=1,\ldots,4$, 
and for both values of $\epsilon$.

\begin{figure}
\begin{center}
\includegraphics[width=9cm,height=6cm]{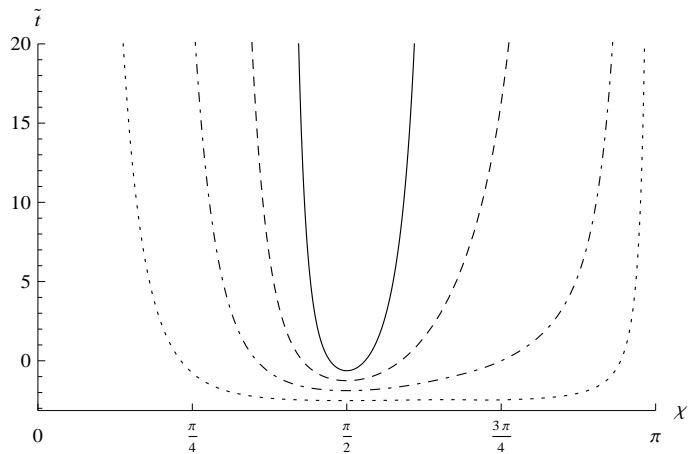}
\caption{Projection onto the $\tilde{t}-\chi$ plane of characteristic curves, 
with $\tilde{t}(\chi=\pi/2)=\eta,\, \eta(\chi=\pi/2)=-k\,\pi,\, k$ = 1/5 (solid
line), 2/5 (dashed line), 3/5 (dash-dotted line), 4/5 (dotted line) initial 
data.}
\label{fig:ttilde_chi}
\end{center}
\end{figure}

The conditions for the variable $\tilde{t}$ are the ones in Proposition 2, once
$F$ is written in terms of $f$ and the components of the initial metric 
(\ref{e:dust_metric}) in this particular case. From Eq.~(\ref{e:def_Gamma}) 
(let us draw reader's attention to the fact that we are using $\eta$ and $\chi$
instead of $t$ and $r$, respectively), and the corresponding values of 
$\tilde{f}$ and $b$, $\tilde{t}$ has to satisfy 
$\left. \partial\tilde{t} / \partial\chi \right|_{\eta=-\pi}=0 \displaystyle$,
i.e., $\displaystyle  \left. \tilde{t} \right|_{\eta = - \pi}$ has to be 
constant. Moreover, 
$\displaystyle \left. \partial\tilde{t} / \partial\eta \right|_{\chi=\pi/2}=1$.
We can choose consistently $\left. \tilde{t} \right|_{\chi=\pi/2} = \eta$ and 
$\left. \tilde{t} \right|_{\eta=-\pi} = -\pi$. The values of $\tilde{t}$ in the
whole domain of parameters can be numerically obtained from the expression of 
$\partial\tilde{t}/\partial\eta$, coefficient of $d\eta$ in 
Eq.~(\ref{e:def_Gamma}) for $\Gamma$, and the condition 
$\left. \tilde{t} \right|_{\chi=\pi/2} = \eta$. In Fig.~\ref{fig:ttilde_chi}
we plot the projection onto the $\tilde{t}-\chi$ plane of the characteristic 
curves, for the values $\tilde{t}(\chi=\pi/2)=\eta$, 
$\eta(\chi=\pi/2)=-k\,\pi$, $k = l/5$, $l=1,\ldots,4$.

Numerically, the metric (\ref{e:dust_metric}) can be written in the new set of 
coordinates $\left\{\tilde{t}, \tilde{r}, \theta, \varphi\right\}$, evaluating 
the corresponding Jacobian matrix from (\ref{e:jacobian}). The hypersurfaces 
$\tilde{t}=$ constant are now maximal.

%%%%%%%%%%%%%%%%%%%%%%%%%%%%%%%%%%%%%%%%%%

\section{Conclusions}
\label{sec:8}
We have proved that any SSST can be locally sliced by a family of maximal 
spacelike hypersurfaces.
 
We have designed a general geometrical method to build up maximal slices in any
SSST by solving three decoupled first order quasi-linear PDEs 
(\ref{e:lambda_r_2}), (\ref{e:int1_2}) and (\ref{e:int1_3}). It has been 
applied in the cases of Minkowski (vacuum) and Friedmann (matter) spacetimes. 
In the first case, a flat spacetime, the equations leading to obtain maximal 
slicings, which are not inertial in general, can be solved analytically. In the
second case, closed Friedmann spacetime, the corresponding equations leading to
obtain maximal slicings are solved numerically.

An interesting by-product for Numerical Relativity of the approach presented in
this paper has to do with the assessment of 3D codes written, as customary, in 
Cartesian coordinates. Let us consider two codes NC1 and NC2 such that only NC1
uses a gauge which is maximal. Hence, the evolution with code NC2 of any 
initial data admitting a spherically symmetric limit could be compared to the 
evolution produced by code NC1, by simply using our procedure to generate a 
SSST satisfying the maximal slicing condition. To emphasize this point we have 
applied our procedure to build up, numerically, maximal slicings in a Friedmann
spacetime. If one takes into account that Oppenheimer-Snyder collapse is one of
the basic test-beds for any numerical code aiming to solve Einstein equations, 
our analysis paves the way to calibrate numerical codes that use formalisms 
that do not incorporate the maximal slicing condition.

We have outlined a method that should be extended to deal with the construction
of constant mean curvature slicings. The results on maximal slicings we have 
presented here can be considered as a step to gain some insights in this 
direction.

Let us make a comment about maximal hypersurfaces in a cosmological scenario. In the seventies, Brill,\cite{Brill} and Brill 
and Flaherty\cite{Brill-Fla} analyzed the existence and uniqueness of maximal hypersurfaces in a cosmological context. Brill concludes 
that {\it closed surface of stationary area (MS) in closed universes are exceptional and scarce};\cite{Brill} 
in particular, Brill finds: {\it Any closed MS in a closed universe which satisfies Einstein's equations with the energy condition is a local maximal surface and is locally 
isolated}.\cite{Brill}  These authors reached these conclusions in analyzing non-static Friedmannian cosmologies, 
for which such a hypersurfaces are identified with instants of cosmic time. In fact,  in these cosmologies, 
the volume functional (relative to the matter flow) might  take a critical value only if the expansion of the 
cosmological observers vanishes, i.e. at a particular cosmological moment, depending on the particular closed universe.  The above conclusions by Brill and Flaherty and the one presented in this paper, i.e., the local existence 
of a maximal slicing in any SSST, are compatible.  In fact, the result by Bill and Flaherty involves the hypothesis of global spatial compacteness meanwhile our result concerns only  local existence, 
and no global topological property has been required.

As we have pointed out from the very beginning, constructing maximal 
slicing are closely related to find a vorticity-free observer without expansion. Such an Eulerian observer is not 
necessary to be geodesic or shear-free, i.e., it is not (generically) a cosmological observer moving with the matter 
content of the universe. Then, as a corollary of the results we have presented, maximal slicings exist locally (and can be 
constructed) in any Robertson-Walker spacetime: they are those slicings  which are associated with Eulerian observers without expansion.

Our statement about the local existence of a maximal slicing in any SSST, and the examples of constructions we have considered, support the physical interest in considering such a local foliations when their global existence is forbidden from topological requirements.

%%%%%%%%%%%%%%%%%%%%%%%%%%%%%%%%%%%%%%%%%%

\section*{Acknowledgements}

I. C.-C. acknowledges support from the Alexander von Humboldt Foundation. 
We wish to thank Dr.~Jos\'e Luis Jaramillo for detailed appreciations and 
interesting comments about this paper and related issues. This work has been 
also supported by the MICINN Grant No. AYA2007-67626-C03-01 and the MICINN-FEDER
project No. FIS2009-07705.

%%%%%%%%%%%%%%%%%%%%%%%%%%%%%%%%%%%%

\section*{References}

\end{document}